# Moderate bending strain induced semiconductor to metal transition in Si nanowires

M. Golam Rabbani[1], Sunil R. Patil[2], M. P. Anantram[1]

[1]Department of Electrical Engineering, University of Washington, Seattle, WA 98195 USA, [2]Department of Physics, College of Engineering, Pune, 411005 MS, India

**Abstract**

Moderate amount of bending strains, ~3% are enough to induce the semiconductor-metal transition in Si nanowires of ~4nm diameter. The influence of bending on silicon nanowires of 1 nm to 4.3 nm diameter is investigated using molecular dynamics and quantum transport simulations. Local strains in nanowires are analyzed along with the effect of bending strain and nanowire diameter on electronic transport and the transmission energy gap. Interestingly, relatively wider nanowires are found to undergo semiconductor-metal transition at relatively lower bending strains. The effect of bending strain on electronic properties is then compared with the conventional way of straining, i.e. uniaxial, which shows that, the bending is much more efficient way of straining to enhance the electronic transport and also to induce the semiconductor-metal transition in experimentally realizable Si nanowires.

## 1. Introduction

Silicon nanowires (SiNWs) have been employed in numerous electronic devices (e.g. nanotransistors [1]–[3], solar cells [4]–[6]), optoelectronic devices [7], sensors [8]–[11], as well as thermoelectric energy conversions [12], [13]. They can be fabricated/synthesized in both top-down and bottom-up approaches, with a wide range of diameters; from a few nanometers to a few hundred nanometers. The sub-10 nm diameter nanowires are particularly interesting, as, unlike the bulk silicon and the wider wires, they can have a direct bandgap [14]–[16]. More interestingly, the bandgap of the narrow SiNW can be efficiently strain engineered [14], [17] to an extent of strain induced reversible direct to indirect bandgap transition [14], [15] and also semiconductor to metal transition [18]. This opens up the possibility and realizations of strain modulated not only electronic but also optoelectronic properties of SiNWs.

The traditional objective of straining semiconductor devices is to enhance their mobility through lattice mismatch and stress-memorization [19] [20]–[22] techniques. For NWs though, bending of a substrate containing the nanowires [23], [24] or deposition of an extra layer of straining materials [25], such as oxide, nitride and carbide [18], on a part or the entire length of the wire are more common. Bottom up synthesized nanowires are also unintentionally [26], [27] or intentionally [28], strained and are found to be stable up to over 40% bending strain [29] and a U-shape bending [30]. In fact it is noted that, as nanowire diameter decreases, the bending strength increases [30]. These studies [29] [30], indicate that narrow SiNWs, in addition to demonstrating quantum mechanical properties, also offer mechanical stability under large stresses.

When it comes to model and simulate the strain effect on electronic properties, there have been numerous attempts, which are usually carried out at the density functional theory (DFT) and classical molecular dynamics (MD) level [17], [31]. However, owing to computationally expensive nature of DFT, the device size is usually limited to few 100s of atoms. Often, the finite size effect is countered by using periodic boundary condition (PBC). Most of the investigations, in this field are therefore limited to uniformly strained material systems. This is largely due to non-uniform strain such as bending does not preserve the crystal symmetry, without which PBC cannot be applied along the length of the nanowire. Alternatively, one may consider long enough NWs, which however, increases the computational demand to the unmanageable level. MD can handle larger system sizes and therefore, is widely used in the study of mechanical [32], [33] and thermal [34], [35] properties of microscale- [36] and nanoscale-devices [37], [38], including silicon nanowires [35], [39]–[50]. However, only reference [50] have studied the effect of bending on electronic properties of nanowires. On the experimental front, it is understood that bending of NWs increases the conductivity in Si as well as in ZnO and CdS [51], [52]. It is to be noted that bending is not limited to only bottom up nanowires, top-down fabricated nanowires can also be bent intentionally [53], [54] or unintentionally [55], [56]. Moreover, top down and bottom up approaches can be combined [57] to get horizontally suspended, well-oriented and size-controlled nanowire arrays. These bending approaches may open up new possibilities, and thus, further investigations on bent nanowires have become crucial.

Therefore, this work investigates the effect of bending on electronic properties of SiNWs. Particularly, the approach involves bent SiNWs optimized with MD, whose output, i.e. atomic co-ordinates are fed to tight binding model to obtain the electronic Hamiltonian. With this Hamiltonian we perform the quantum transport calculation within the non-equilibrium Green's function [58] formulation. We also present the comparative results of bent nanowires with that of the uniaxially strained nanowires. The paper is organized as follows, the section 2 describes the simulation methodology followed by section 3, which includes strain analysis, results and discussion on the effect of bending on electronic properties of NWs, and the section 4 concludes the work with the conclusion section.

## 2. Methodology

Six different cross-sectional area, <110> directed SiNWs are studied, as sub-10nm diameter nanowires grow mainly along the <110> direction [27], [59]–[63]. The average diameters of the nanowires are 1.0 nm, 1.7 nm, 2.3 nm, 3.1 nm, 3.6 nm and 4.3 nm, which are labeled here as <110>2d, <110>3d, <110>4d, <110>5d, <110>6d and <110>7d, respectively. The total length of each nanowire is 19.922 nm with 103 atomic layers. The initial nanowire structures were generated by the repetition of DFT optimized silicon unit cells [14]. The MD is performed with the Large-scale Atomic/Molecular Massively Parallel Simulator (LAMMPS) [64]. The Tersoff [65], [66] many-body bond-order reactive potential is used to describe the Si-Si interactions. The nanowire structure is first optimized by using energy minimization, with no velocity or force restrictions on any atoms. We follow bending procedure, which was used for bending the silicon carbide nanowires [32], where the Si atoms are subdivided into three groups. To apply bending stress to the middle group, the atoms in the left (right) group are rotated counter clockwise (clockwise) in the yz plane (Supplementary Information Figure 1(b) top) by an angle of 0.05 degree. Then they are kept frozen while energy minimization is done to the middle group of atoms. After the minimization, the atomic positions of all the atoms are saved, and the process is repeated for the next increment of the rotation angle. Similar approach is used for uniaxial strain, by shifting the end atoms, instead of rotating them. Since the left and right groups of atoms are kept fixed during the bending or uniaxial straining and energy minimization process, the atoms in the middle group, which are close to atoms in the edge groups, undergo artificial deformations. To avoid this, 19 layers of atoms from each end are cropped off. Next, the remaining 65 layers are hydrogen passivated with the GaussView [67], and then were used for tight binding and quantum transport calculations. We employ the $sp^3d^5s^*$ tight binding (TB) method with parameters from Jancu [68] to form electronic Hamiltonian matrix. Without considering spin-orbit coupling in $sp^3d^5s^*$ TB,

Hamiltonian size for each silicon atom is 10x10, and that for each hydrogen atom is 1x1. Thus, the Hamiltonian size of each layer is between 128x128 (<110>2d) and 1498x1498 (<110>7d). The Hamiltonian formed by TB is used to calculate the transmission through the nanowire by using the non-equilibrium Green's function (NEGF)[58] formalism. For ballistic transport, the electronic transmission in a semiconductor is zero in the energy bandgap, therefore, transmission can be used to estimate the energy bandgap.

## 3. Results and Discussion

### 3.1 Strain analysis

In this subsection, we analyze the strains in the nanowires. Figure 1 shows three bent nanowires, (a) <110>3d, (b) <110>5d, and (c) <110>7d, with their local atomic strain along the length (z) direction visualized as color variation. To obtain this strain configuration, the left and right atom groups of each nanowire was rotated by 15º in the vertical (yz) plane. The top side of each nanowire is under tension while the bottom side is under compression. The corner atoms are under the largest amount of strain due to the rotation. Also for a fixed amount of rotation, as expected, the range of strain variation (from most compressive to most tensile) increases with the nanowire diameter. In addition, the largest tensile strain magnitude is always higher than the largest compressive strain magnitude, for example, +3.7% vs -2.3% for nanowire <110>3d. This is due to the atomic potentials resisting compression more than tension because as the atoms are brought closer to each other beyond the no-strain minimum energy position the energy increases exponentially.

Figure 2, therefore, plots all the three strain ranges (x-range (a), y-range (b), and z-range (c)) as a function of the nanowire diameter for different amount of end-atom-group rotations. The ranges are almost linearly proportion to the nanowire diameter. The y-range values are slightly higher than the x-range values as diameter along y is wider. Due to bending along the length, z-range values are the largest.

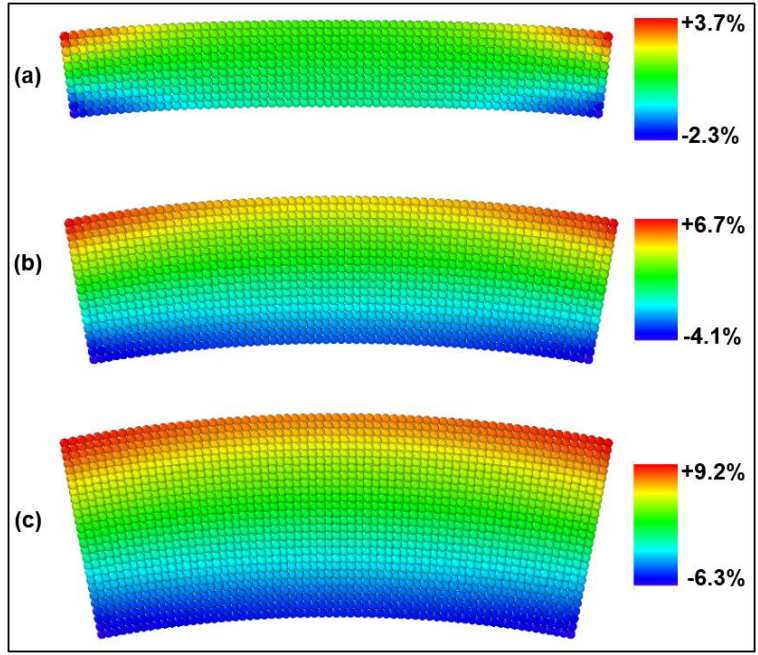

Figure 1. Local strain along the **z-direction** (**length**) in 3 different nanowires as a color plot; (a) <110>3d, (b) <110>5d, (c) <110>7d.

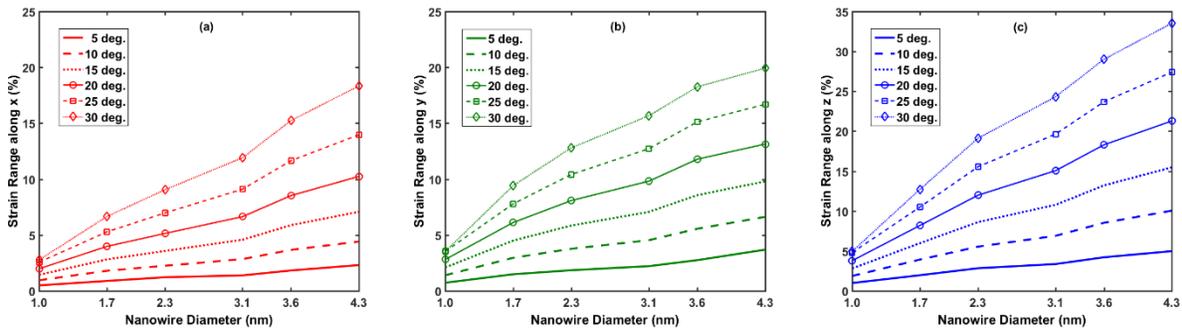

Figure 2. Ranges of strain (x-range (a), y-range, (b) and z-range (c)) as functions of nanowire diameter for different amount of end-atom-group rotations. x and y directions are in the cross-sectional plane while z is along the length. x dimension is smaller than y dimension. For each case, bottom curve is for smallest rotation (5 deg.) and the rotation increases upwards.

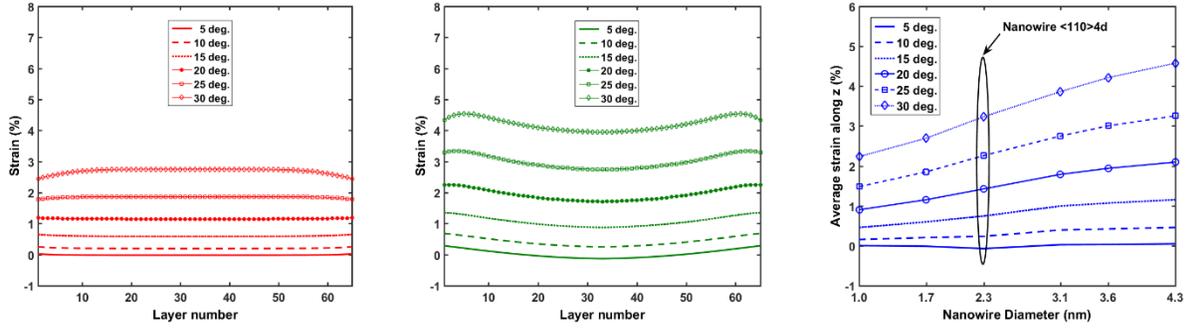

Figure 3. Layer to layer average strain along length: (a) <110>3d, and (b) <110>6d. (c) Average strain along length as a function of diameter for all six nanowires. Nanowire <110>4d has been pointed out.

Figure 3(a) and b) plots the layer to layer strain along the length, as defined in equation (1), for two different nanowires - (a) <110>3d and (b) <110>6d. We have already seen in Figure 2(c) that the range of variation of z-strain over all the atoms in a nanowire can be quite large, mainly because of the corner atoms being under large strain of opposite polarity. However, the layer to layer strain variation is actually smaller (Figure 3(a) and (b)). As variation is small, it is possible to find a single quantity, representing an equivalent average strain, for each of the curves in Figure 3. This average number is plotted in Figure 3(c) for all the bent nanowires considered in this study. For the same end-atom rotation, the average strain increases linearly with the diameter, like the strain ranges. The average strain value is used for studying correlation with electronic properties and comparison with uniaxially strained nanowires, for which nominal strain value along length is used.

**3.2 Electronic transmission**

In bent nanowires, the bending strain compresses or extends the nanowire locally (Figure 1), although on an average, the nanowire is under tensile strain along the length (Figure 3(c)). The strain (bending or uniaxial) changes the electronic band profile of the nanowire [69]. Due to the absence of crystal symmetry under bending, one cannot determine the energy band structure using the Bloch's theorem. We rather calculate the electronic transmission (Supplementary Information Eq. (5)) in the strained nanowires as a function of energy, which has information about the energy gap as well as transmission probability.

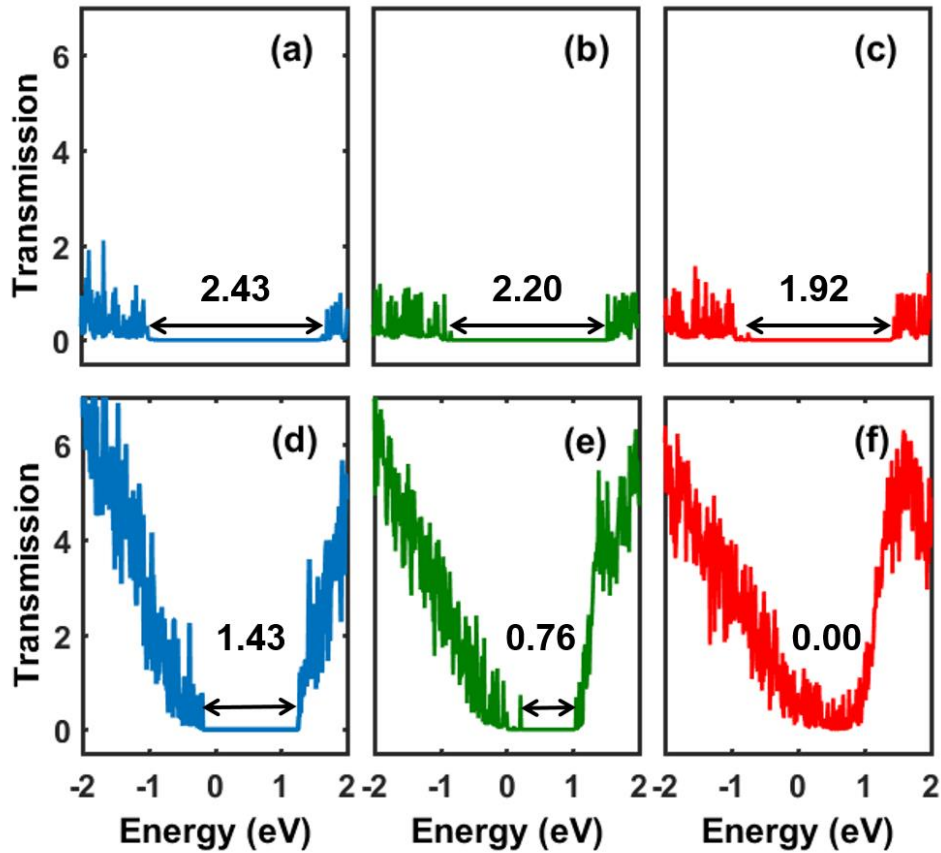

Figure 4. Electronic transmission through two nanowires as a function of energy for different bending angles. Narrowest <110>2d nanowire with no rotation (a), with 15-degree rotation (b), with 30-degree rotation (c). Widest <110>7d nanowire with no rotation (d), with 15-degree rotation (e), 30-degree rotation (f).

Figure 4 shows the electronic transmission through the nanowires as a function of energy for, respectively, zero, 15-degree and 30-degree end-atom-group rotations: (a, b, c) <110>2d and (d, e, f) <110>7d. We notice three important trends in the plots. First, a narrower nanowire has a lower transmission value. The reason is a narrower nanowire has a fewer number of electronic modes, and, at any energy, transmission is proportional to the number of modes. Second, the transmission gap - the continuous energy range over which the transmission is zero - in the nanowire is larger than the bulk silicon bandgap. As a result of quantum confinement decreasing with increasing NW diameter, the transmission gap decreases as the nanowire diameter increases. The bandgap values, calculated by assuming the entire length of the bent nanowire as a single unit cell, also match with the corresponding transmission gap. In addition, it is to be noticed that the gaps in the unstrained nanowires, 2.43 eV (<110>2d) and 1.43 eV (<110>7d), are consistent with those obtained with DFT [14] and DFTB [16]. Third and most importantly, the gap in each nanowire decreases as the amount of bending strain (that is, average strain along length) increases. This behavior is also consistent with the results in [14], which predicts that for an increase in uniaxial tensile strain in the nanowire, its bandgap decreases.

We can get a more comprehensive overview by studying the transmission energy gap as a function of the nanowire diameter with the angle of end atom group rotation as a parameter. Such characteristics are plotted in Figure 5 for no rotation and six different rotations (5, 10, 15, 20, 25, and 30 degree) conditions (solid lines). For small end-atom rotations (top curves), as the nanowire diameter decreases from the largest diameter, 4.3 nm, the gap increases gradually. This increase, however, becomes sharper as the diameter decreases below 1.7 nm. For large rotations, the gap increases almost linearly as the diameter decreases. It is interesting to note that for the two widest wires, diameters 3.6 nm and 4.3 nm, the gap completely disappears for large strains ~3%, as the strain range is higher for the wider wires. For large bending, the local strains (both tensile and compression) in some parts of the wider wires are very high (and of opposite direction). These large local strains represent a huge departure from the perfect crystal structure and introduce many electronic modes in the band gap, thereby making the gap totally disappear (Figure 4(f)). This particular phenomena has been reported in relatively thinner core-shell Si-SiC NWs and attributed to the relatively huge compressive strain ~9% [18]. It has been shown experimentally that bending increases nanowire conductivity [51], [52], so bandgap decrease and conductivity increase can be correlated.

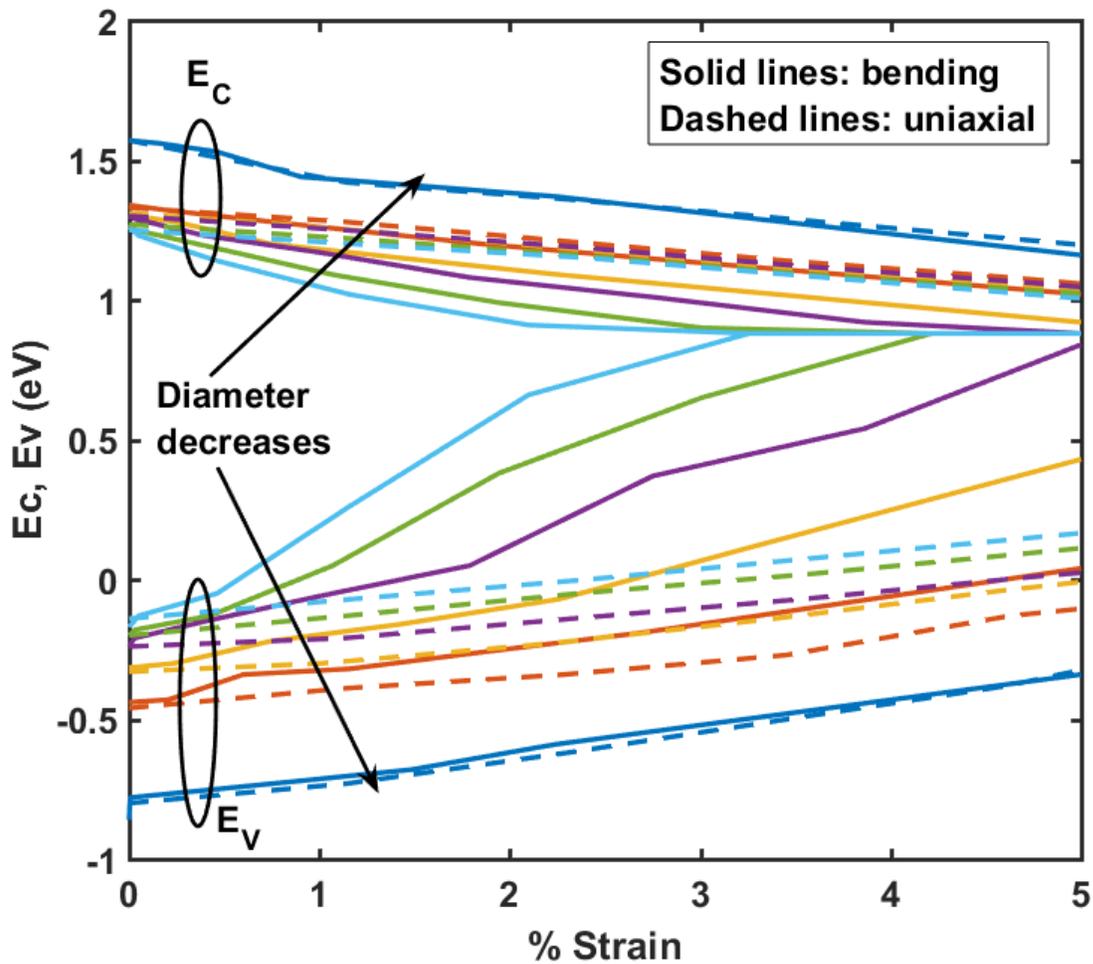

Figure 5. $E_C$ and $E_V$ as a function of % strain. Solid: bent, dashed: uniaxial.

### 3.3 Bent nanowire vs. uniaxially strained nanowire

Figure 5 also shows the energy gap in electronic transmission as a function of strain for six nanowires uniaxially strained cases (dashed lines). The gap decreases linearly with the strain for both the method of straining, which agrees with uniaxially strained nanowires studied with DFT [14]. However, for the same diameter and the same average strain, the gap is smaller and its decrease rate is faster in the bent nanowires. The larger decrease rate also means that the strain modulated band gap will respond to wider electromagnetic spectrum. This clearly is an advantage of the bent nanowire over the uniaxially strained nanowire.

It is also to be noted that, for narrow nanowires, both $E_C$ and $E_V$ change at a similar rate, but as nanowire diameter increases, the rate of change of $E_V$ becomes faster, indicating that the bending affects the valence band maxima significantly than that of the conduction band which is also consistent with [18], where similar observations are noted under relatively large compressive strains of ~9%. Similar trends are seen in bi-axially strained silicon and has been attributed to valence band valley splitting in the effective mass approach [69]. It is intricate, which requires additional investigations, to pin point the particular phenomenon leading to the significant upward shift in valence band edge due to the complicated bending strain profile.

We also would like to note here that, although the maximum value of experimentally observed uniaxial strain is about 5% [25], nanowires have been found to withstand large (40%) bending strain [29]. In another study, U-shape bent 40 nm diameter silicon nanowire was stable up to 11.5% strain [30]. So considering the $E_C$ and $E_V$ trends vs strain in Figure 5 and the amount of experimentally observed bending strain can be easily achieved experimentally in SiNWs.

## 4. Conclusion

Electronic transport properties SiNWs of diameters from 1 nm to 4.3 nm under bending strain are simulated using MD for structural optimization and NEGF for transport calculations. It is observed that for the same amount of bending, strain range increases with nanowire diameter. The larger strains directly translate into smaller energy gaps in electronic transmission for wider wires. The electronic transport in bent SiNWs are also compared with that of uniaxially strained. Bending is found to be a more efficient way as compared to uniaxial strain, of enhancing the electronic transport and reducing transmission gap, such that SiNWs undergo semiconductor to metal transition at relatively lower strain values. This makes the bending as an efficient straining mechanism over uniaxial in nanowires in respect to conductance enhancements well as modulation of energy gaps to exploit wider electromagnetic spectrum.